\documentclass[aps,showpacs,twocolumn]{revtex4}
\usepackage{graphicx}
\begin{document}
\title{Nature of chiral phase transition in QED$_3$ at zero density}
\author{Hong-tao Feng$^{1,5}$\footnote{Email: fenght@seu.edu.cn}, Jian-Feng Li$^{2,5}$, Yuan-mei Shi$^{3,4,5}$, and Hong-shi Zong$^{4,5,6}$\footnote{Email:zonghs@chenwang.nju.edu.cn}}
\address{$^1$Department of Physics, Southeast University, Nanjing,  211189, China}
\address{$^2$College of Mathematics and Physics, Nantong University, Nantong 226019, China}
\address{$^3$Department of Physics, Nanjing Xiaozhuang College, Nanjing 211171, China}
\address{$^4$Department of Physics, Nanjing University, Nanjing, 210093, China}
\address{$^5$ State Key Laboratory of Theoretical Physics, Institute of Theoretical Physics, CAS, Beijing 100190, People¡¯s Republic of China}
\address{$^6$ Joint Center for Particle, Nuclear Physics and Cosmology, Nanjing 210093, China}
\begin{abstract}
Based on the feature of chiral susceptibility and thermal
susceptibility at finite temperature, the nature of chiral phase
transition around the critical number of fermion flavors ($N_c$) and
the critical temperature ($T_c$) at a fixed fermion flavors number in
massless QED$_3$ are investigated. It is showed that, at finite
temperature the system exhibits a second-order phase transition at
$N_c$ or $T_c$ and each of the estimated critical exponents is less
than 1, while it reveals a higher-order continuous phase transition
around $N_c$ at zero temperature.

\bigskip

\end{abstract}
\maketitle

 \maketitle
\section{introduction}
The study of chiral phase transition (CPT) in (2+1)-dimensional
quantum electrodynamics (QED$_3$) has been an active subject for 30
years since Appelquist \emph{et al}. found that CPT occurs when the
flavor of massless fermions reaches a critical number $N_c$
\cite{a1}. They arrived at this conclusion by analytically and
numerically solving the Dyson-Schwinger equation (DSE) for the
fermion self-energy in the lowest order approximation where the
involved one-loop boson polarization is obtained by the free form of
the fermion propagator. To indicate the value of $N_c$, D. Nash
adopted an improved scheme and gave a larger $N_c$ \cite{a2}. Later,
several groups investigated the dependence of chiral symmetry
breaking on $N$ and some groups doubted the existence of $N_c$
\cite{a3,a4}. This question was answered by P. Maris \emph{et al},
who used the coupled DSEs for the photon and fermion propagator to
investigate the influence of the full vacuum polarization and vertex
function on the fermion propagator and they found that the critical
number of fermion flavors for dynamical mass generation of massless
QED$_3$ lies between 3 and 4 \cite{a5,a5a,a6}.

Nevertheless, what is the order of CPT around $N_c$ might be an
interesting question. To reveal that, the authors of \cite{a1}
studied  the light scalar degrees of freedom and the order parameter
of CPT near $N_c$ and found that the phase transition is not
second-order and is also unlike conventional first-order transition
\cite{a6a}. In addition, the results from Cornwall-Jackiw-Tomboulis
effective potential also gave the same conclusion \cite{a6b}.
Although the above reveals the characteristic CPT, it is interesting
to adopt an alternative method to reanalyze the nature of this phase
transition and see whether it is consistent with those results.

At finite temperature, the value of $N_c$ should also vary and
chiral symmetry restores as the temperature increases at a fixed
$N(<N_c)$. In this case, the fermion propagator at finite
temperature $T$ can be written as
\begin{equation}\label{FPT}
S^{-1}(T, P)=i\vec\gamma\cdot\vec PA_\|(P^2)+i\varpi_n
\gamma_3A_3(P^2)+B(P^2),
\end{equation}
where $\varpi_n=(2n+1) \pi T$ and $A,~B$ are the fermion
wave-renormalization factor and self-energy, respectively. Adopting
the lowest-order approximation of DSE and using Eq. (\ref{FPT}), Dorey
investigated the CPT of QED$_3$ at finite temperature and showed
that QED$_3$ with dynamical chiral symmetry breaking (DCSB)
undergoes CPT into chiral symmetric phase when the temperature
reaches a critical value $T_c$ and the corresponding $N_c$
decreases with the increasing temperature \cite{a7}.

The above conclusion holds in massless QED$_3$. Then, another
natural question may be raised: how does one chart the phase diagram
of thermal QED$_3$ around $T_c$ and whether or not the nature of CPT
around $N_c$ at finite temperature is the same as that at zero
temperature. At the involved temperature, since the external fields
are screened by thermal excitations and the boson gains a nonzero
mass, the feature of CPT at $N_c$ might be changed. However, as far
as we know, the nature of CPT at $N_c$ in thermal QED$_3$ has not
been reported in the existing literature. Therefore, it is very
interesting to study this problem.

In recent years, some works in lattice QCD \cite{a8,a9,a10} showed
that the peak of chiral susceptibility should be an essential
characteristic of CPT. Later, based on techniques of continuum field
theory, several groups \cite{a11,a12,a13,a14,a15,a16} also reached
the same conclusion. Thus, chiral susceptibility is competent for
studying the feature of phase transition in this nonperturbative
system. Mealwhile, the thermal susceptibility give other ideal
parameter to investigate the characteristic of CPT at finite
temperature\cite{a16a}. In this paper, we shall adopt the chiral and
thermal susceptibilities to study the nature of chiral phase
transition of QED$_3$ at finite temperature.

\section{formalism for chiral susceptibility}
The Lagrangian of QED$_3$ involving $N$ fermion flavors of
$4\times1$ spinor reads
\begin{equation}\label{LQED}
    \mathcal{L}=\sum^N_{j=0}\bar\psi_j(\not\!\partial+i\mathrm{e}\not\!\!A-m)\psi_j+\frac{1}{4}F^2_{\sigma\nu}+\frac{1}{2\xi}(\partial_\rho A_\rho)^2.
\end{equation}
In the absence of the mass term $m\bar\psi\psi$, QED$_3$ has chiral
symmetry. There are several equivalent choices of the order
parameter for chiral symmetry breaking; here we use the fermion
chiral condensate
\begin{equation}\label{DPC}
\langle\bar\psi\psi\rangle_m=\int\frac{\mathrm{d}^3p}{(2\pi)^3}\mathrm{Tr}[S(m,p)],
\end{equation}
where $S$ is the dressed fermion propagator and $Tr$ denotes trace
operation over Dirac indices of the fermion propagator. Based on
Lorentz structure analysis, the involved massive/massless fermion
propagator can be written as
\begin{eqnarray}
    &&S^{-1}(m,p)=i\gamma\cdot pE(p^2)+F(p^2),\\
    &&S^{-1}(p)\equiv S^{-1}(0,p)=i\gamma\cdot pA(p^2)+B(p^2).
    \label{S0}
\end{eqnarray}
In the high energy limit, the fermion propagator reduces to the free
one, i.e., $ S_0^{-1}(p)=i\gamma\cdot p$ in the chiral limit and
$S_0^{-1}(m,p)=i\gamma\cdot p+m$ beyond the chiral limit. From this it
can be seen that, with a small fermion mass $m$, the integral in Eq.
(\ref{DPC}) is divergent. In this case we should employ a
renormalization procedure to deal with this divergence. A natural
approach is to subtract the condensate of the free fermion field
from the above value. That is to say, we define the renormalized
fermion chiral condensate by
\begin{equation}\label{renormsus}
\langle\bar\psi\psi\rangle\equiv\langle\bar\psi\psi\rangle_m-\langle\bar\psi\psi\rangle_{mf},
\end{equation}
where $\langle\bar\psi\psi\rangle_{mf}$ is the condensate of the
free fermion gas. Below, we shall determine the transition point via
the maximum of chiral susceptibility
$\frac{\partial{\langle\bar\psi\psi\rangle}}{\partial{m}}$ (see,
e.g., Refs. \cite{a8,b1}) which is defined as\cite{a11}
\begin{equation}\label{DCF}
\chi^c=\left.\frac{\partial \langle\bar \psi\psi\rangle}{\partial
m}\right|_{m\rightarrow0}.
\end{equation}
This equation indicates that the chiral susceptibility measures the
response of the chiral condensate (the order parameter) to an
infinitesimal change of the fermion mass responsible for explicit
breaking of chiral symmetry. Note here that we evaluate the chiral
susceptibility and fermion chiral condensate in the chiral limit.

From Eqs. (\ref{DPC})-(\ref{S0}), we immediately arrive at the chiral
susceptibility of QED$_3$ in chiral limit
\begin{equation}\label{chic}
  \chi^c
  =4N\int\frac{\mathrm{d}^3p}{(2\pi)^3}\left\{\frac{p^2A^2D-2p^2ABC-B^2D}{[p^2A^2+B^2]^2}-\frac{1}{p^2}\right\},
\end{equation}
where
\begin{equation}\label{CD}
C(p^2)=\left.\frac{\partial E(p^2)}{\partial
m}\right|_{m\rightarrow0},~D(p^2)=\left.\frac{\partial
F(p^2)}{\partial m}\right|_{m\rightarrow0}.
\end{equation}

\section{Zero temperature}

 The next task is to obtain the four functions $A,~B,~C,~D$.
These functions can be obtained by solving the DSE for the massive
fermion propagator,
\begin{eqnarray}
S^{-1}(m,p)&=&S^{-1}_{0}(m,p)+\int\frac{\mathrm{d}^{3}k}{(2\pi)^{3}}\times\nonumber\\&&[\gamma_{\sigma}S(m,k)\Gamma_{\nu}(m;p,k)D_{\sigma\nu}(m,q)],
\label{eq2}
\end{eqnarray}
where $\Gamma_{\nu}(m;p,k)$ is the full fermion-photon vertex and
$q=p-k$. The coupling constant $\alpha=e^2$ has dimension one, and
provides us with a mass scale. For simplicity, in this paper
temperature, mass and momentum are all measured in unit of $\alpha$,
namely, we choose a kind of natural units in which $\alpha= 1$. Form
Eq. (\ref{S0}) and Eq. (\ref{eq2}), we obtain the equation satisfied
by $E(p^2)$ and $F(p^2)$
\begin{eqnarray}
E(p^{2})&=&1-\frac{1}{4p^2}\int\frac{\mathrm{d}^{3}k}{(2\pi)^{3}}Tr[i(\gamma p)\gamma_{\sigma}S(m,k)\times\nonumber\\
&&~~~~~~~~~~~~~~\Gamma_{\nu}(m;p,k)D_{\sigma\nu}(m,q)],\\
F(p^{2})&=&\frac{1}{4}\int\frac{\mathrm{d}^{3}k}{(2\pi)^{3}}\times\nonumber\\&&Tr\left[\gamma_{\sigma}S(m,k)\Gamma_{\nu}(m;p,k)D_{\sigma\nu}(m,q)\right].
\end{eqnarray}
Another involved function $D_{\sigma\nu}(q)$ is the full gauge boson
propagator which is given by \cite{a14}
\begin{equation}
D_{\sigma\nu}(m,q)=\frac{\delta_{\sigma\nu}-q_{\sigma}q_{\nu}/q^{2}}{q^{2}[1+\Pi(m,q^{2})]}+\xi\frac{q_\sigma
q_\nu}{q^4},
\end{equation}
where $\xi$ is the gauge parameter and $\Pi(q^{2})$ is the vacuum
polarization for the gauge boson which is satisfied by the
polarization tensor for gauge boson and reads
\begin{equation}
\Pi_{\sigma\nu}(m,q^{2})=-\int\frac{d^{3}k}{(2\pi)^{3}}Tr\left[S(m,k)\gamma_{\sigma}S(m,q+k)\Gamma_{\nu}(m,p,k)\right].
\end{equation}

Using the relation between the vacuum polarization $\Pi(m,q^{2})$
and $\Pi_{\sigma\nu}(q^{2})$,
\begin{equation}
\Pi_{\sigma\nu}(m,q^{2})=(q^2\delta_{\sigma\nu}-q_{\sigma}
q_{\nu})\Pi(m,q^{2}),
\end{equation}
we can obtain an equation for $\Pi(q^2)$ which has ultraviolet
divergence. Fortunately, it is present only in the longitudinal part
and is proportional to $\delta_{\sigma\nu}$. We can remove the
divergence by the projection operator
\begin{equation}
\mathcal{P}_{\sigma\nu}=\delta_{\sigma\nu}-3\frac{q_{\sigma}q_{\nu}}{q^2},
\end{equation}
and obtain a finite vacuum polarization\cite{a15}.

Finally, we choose to work in the Landau gauge, since the Landau
gauge is the most convenient and commonly used one. Once the
fermion-boson vertex is known, we immediately obtain truncated DSEs
for the propagators of the fermion and the gauge boson and then the
chiral susceptibility near $N_c$ is obtained. Of course, just as
mentioned in Ref. \cite{b3}, $N_c$ occurs only in homogeneous
system, i.e., all the involved functions in this issue for the
fermion and boson propagators should satisfy homogeneity degrees.

\subsection{Rainbow approximation} The simplest and most commonly
used truncated scheme for the DSEs is the rainbow approximation,
\begin{equation}
\Gamma_\nu\rightarrow\gamma_\nu,
\end{equation}
since it gives us rainbow diagrams in the fermion DSE and ladder
diagrams in the Bethe-Salpeter equation for the fermion-antifermion
bound state amplitude. In the framework of this approximation, the
coupled equation for massive fermion propagator reduces to
\begin{equation}
S^{-1}(m,p)=S^{-1}_{0}(m,p)+\int\frac{\mathrm{d}^{3}k}{(2\pi)^{3}}\gamma_{\sigma}S(m,k)\gamma_{\nu}D_{\sigma\nu}(m,q).
\label{eq2}
\end{equation}
From Eq. (\ref{S0}) and Eq. (\ref{eq2}), we obtain the equation
satisfied by $E(p^2)$ and $F(p^2)$
\begin{eqnarray}
E(p^{2})&=&1-\frac{1}{4p^2}\int\frac{\mathrm{d}^{3}k}{(2\pi)^{3}}Tr[i(\gamma
p)\gamma_{\sigma}S(m,k)\gamma_{\nu}D_{\sigma\nu}(m,q)],\nonumber\\\\
F(p^{2})&=&\frac{1}{4}\int\frac{\mathrm{d}^{3}k}{(2\pi)^{3}}Tr\left[\gamma_{\sigma}S(m,k)\gamma_{\nu}D_{\sigma\nu}(m,q)\right],
\end{eqnarray}
In order to obtain these two functions, we start from the
propagators with massive fermion. From the above two equations and
some tricks proposed in Ref. \cite{b4}, we obtain the three coupled
equations for $E(p^2)$, $F(p^2)$ and $\Pi(m,q^2)$,
\begin{eqnarray}\label{RE11}
E(p^2)&=&1+\frac{2}{p^2}\int\frac{\mathrm{d}^{3}k}{(2\pi)^3}\frac{E(k^2)(pq)(kq)/(q^2)^2}{G(k^2)[1+\Pi(m,q^2)]},\\
\label{RFp}
F(p^2)&=&m+2\int\frac{\mathrm{d}^{3}k}{(2\pi)^3}\frac{F(k^2)/q^2}{G(k^2)[1+\Pi(m,q^2)]},\\
\label{Rpi1}
\Pi(m,q^2)&=&\frac{2N}{q^2}\int\frac{\mathrm{d}^{3}k}{(2\pi)^3}\frac{E(k^2)E(p^2)}{G(k^2)G(p^2)}\times\nonumber\\&&[2k^2-4(k\cdot
q)-6(k\cdot q)^2/q^2],
\end{eqnarray}
with $G(k^2)=E^2(k^2)k^2+F^2(k^2)$.

 Adopting Eqs. (\ref{CD}) and (\ref{RE11}-\ref{Rpi1}) and setting
$\Pi'(q^2)=\frac{\partial \Pi(m,q^2)}{\partial m}|_{m\rightarrow0}$,
we get the coupled equations
 for $C(p^2),~D(p^2)$ and $\Pi'(q^2)$,
\begin{eqnarray}
C(p^2)&=&\frac{2}{p^2}\int\frac{\mathrm{d}^3k}{(2\pi)^3}\frac{(p\cdot q)(k\cdot q)C_1/(q^2)^2}{H^2(k^2)\left[1+\Pi(q^2)\right]^2}, \\
D(p^2) &=& 1+2\int\frac{\mathrm{d}^3k}{(2\pi)^3}\frac{D_1/q^2}{H^2(k^2)\left[1+\Pi(q^2)\right]^2},\\
\Pi'(q^2)&=&\frac{2N}{q^2}\int\frac{\mathrm{d}^3k}{(2\pi)^3}\frac{\left[2k^2-4(k\cdot
q)-6(k\cdot q)^2/q^2\right]\Pi'_1}{H^2(k^2)H^2(p^2)},\nonumber\\
\end{eqnarray}
with $H(k^2)=A^2(k^2)k^2+B^2(k^2)$ and
\begin{widetext}
\begin{eqnarray*}
C_1&\equiv&[B^2(k^2)C(k^2)-A^2(k^2)C(k^2)k^2-2A(k^2)B(k^2)D(k^2)]\left[1+\Pi(q^2)\right]-A(k^2)H(k^2)\Pi'(q^2), \\
D_1&\equiv&[A(k^2)D(k^2)k^2-B^2(k^2)D(k^2)-2A(k^2)B(k^2)C(k^2)k^2]\left[1+\Pi(q^2)\right]-B(k^2)H(k^2)\Pi'(q^2), \\
\Pi'_1&\equiv&\left[A(p^2)C(k^2)+A(k^2)C(p^2)\right]H(k^2)H(p^2)-2A(k^2)A(p^2)\left[A(k^2)C(k^2)k^2+B(k^2)D(k^2)\right]H(p^2)\nonumber\\
&&-2A(k^2)A(p^2)\left[A(p^2)C(p^2)p^2+B(p^2)D(p^2)\right]H(k^2),
\end{eqnarray*}
\end{widetext}
where $A,~B,~\Pi$ are obtained by Eqs. (\ref{RE11}-\ref{Rpi1}) at
$m=0$. By application of iterative methods, we can obtain
$A,~B,~\Pi$ and the above functions for the scalar vertex.
\subsection{Improved scheme for DSE} To improve the truncated scheme
for DSE, there are several attempts to determine the functional form
for the full fermion-gauge-boson vertex \cite{b4a,b4b,b4c,b4d}, but
none of them completely resolve the problem. However, the
Ward-Takahashi identity (WTI)
\begin{equation}
(p-k)_\nu\Gamma_{\nu}(m;p,k)=S^{-1}(m,p)-S^{-1}(m,k),
\end{equation}
provides us an effectual tool to obtain a reasonable ansatze for the
full vertex \cite{b4a}.  The portion of the dressed vertex which is
free of kinematic singularities, i.e. BC vertex, can be written as,
\begin{eqnarray}
  \Gamma_\nu(m,p,k) &=& \frac{E(p^2)+E(k^2)}{2}\gamma_\nu +\frac{F(p^2)-F(k^2)}{p^2-k^2}(p+k)_\nu\nonumber\\
  &&+(\not\!p+\not\!k)\frac{E(p^2)-E(k^2)}{2(p^2-k^2)}(p+k)_\nu.
\end{eqnarray}
Since the numerical results obtained using the first part of the
vertex coincide very well with earlier investigations \cite{a6,a13},
we choose this one as a reasonable ansatze
\begin{equation}\label{BCM1}
    \Gamma_\nu^{BC_1}(m;p,k)\doteq\frac{1}{2}\left[E(p^2)+E(k^2)\right]\gamma_\nu
\end{equation}
to be used in our calculation. Following the procedure in rainbow
approximation, we also obtain the three coupled equations for
$E(p^2),~F(p^2)$ and $\Pi(m,q^2)$ in the improved truncated scheme
for DSEs,
\begin{widetext}
\begin{eqnarray}\label{E11}
&&E(p^2)=1+\int\frac{\mathrm{d}^{3}k}{(2\pi)^3}\frac{E(k^2)[E(p^2)+E(k^2)](pq)(kq)/(q^2)^2}{p^2G(k^2)[1+\Pi(m,q^2)]},\\
\label{Fp}
&&F(p^2)=m+\int\frac{\mathrm{d}^{3}k}{(2\pi)^3}\frac{[E(p^2)+E(k^2)]F(k^2)/q^2}{G(k^2)[1+\Pi(m,q^2)]},\\
\label{pi1}
&&\Pi(m,q^2)=\frac{N}{q^2}\int\frac{\mathrm{d}^{3}k}{(2\pi)^3}\frac{E(k^2)E(p^2)[E(p^2)+E(k^2)]}{G(k^2)G(p^2)}[2k^2-4(k\cdot
q)-6(k\cdot q)^2/q^2],
\end{eqnarray}
and the corresponding unknown functions for
$C(p^2),~D(p^2),~\Pi'(q^2)$ are,
\begin{eqnarray*}
C(p^2)&=&\frac{1}{p^2}\int\frac{\mathrm{d}^{3}k}{(2\pi)^3}\frac{(pq)(kq)/q^2}{[1+\Pi(q^2)]^2}\left\{\frac{[C_1-C_2][1+\Pi(q^2)]}{H^2(k^2)}-\frac{A(k^2)[A(p^2)+A(k^2)]\Pi'(q^2)}{H(k^2)}\right\},\\
D(p^2)&=&1+\int\frac{\mathrm{d}^{3}k}{(2\pi)^3}\frac{1}{[1+\Pi(q^2)]^2}\left\{\frac{[D_1-D_2][1+\Pi(q^2)]}{H^2(k^2)}-\frac{B(k^2)[A(p^2)+A(k^2)]\Pi'(q^2)}{H(k^2)}\right\}, \\
\Pi'(q^2)&=&
\frac{N}{q^2}\int\frac{\mathrm{d}^{3}k}{(2\pi)^3}\frac{\Pi'_1\Pi'_2-2\Pi'_3\Pi'_4}{H^2(k^2)H^2(p^2)}[2k^2-4(k\cdot
q)-6(k\cdot q)^2/q^2],
\end{eqnarray*}
with
\begin{eqnarray*}
&&C_1\equiv\{2A(k^2)C(k^2)+A(p^2)C(k^2)+A(k^2)C(p^2)\}H(k^2), \\
&&C_2\equiv 2A(k^2)[A(p^2)+A(k^2)][A(k^2)C(k^2)k^2+B(k^2)D(k^2)],\\
&&D_1\equiv\{D(k^2)[A(k^2)+A(p^2)]+B(k^2)[C(k^2)+C(p^2)]\}H(k^2), \\
&&D_2\equiv 2B(k^2)[A(p^2)+A(k^2)][A(k^2)C(k^2)k^2+B(k^2)D(k^2)], \\
&&\Pi'_1\equiv [A(k^2)C(p^2)+A(p^2)C(k^2)][A(k^2)+A(p^2)]+A(k^2)A(p^2)[C(k^2)+C(p^2)], \\
&&\Pi'_2\equiv H(k^2)H(p^2), \\
&&\Pi'_3\equiv A(k^2)A(p^2)[A(p^2)+A(k^2)], \\
&&\Pi'_4\equiv[A(k^2)C(k^2)k^2+B(k^2)D(k^2)]H(p^2)+[A(p^2)C(p^2)p^2+B(p^2)D(p^2)]H(k^2),
\end{eqnarray*}
\end{widetext} where $A,~B,~\Pi$ are obtained by Eqs. (\ref{E11}-\ref{pi1}) in
the chiral limit.

\subsection{Chiral susceptibility around  $N_c$} 
\begin{figure}[htp!]
  \includegraphics[width=0.22\textwidth]{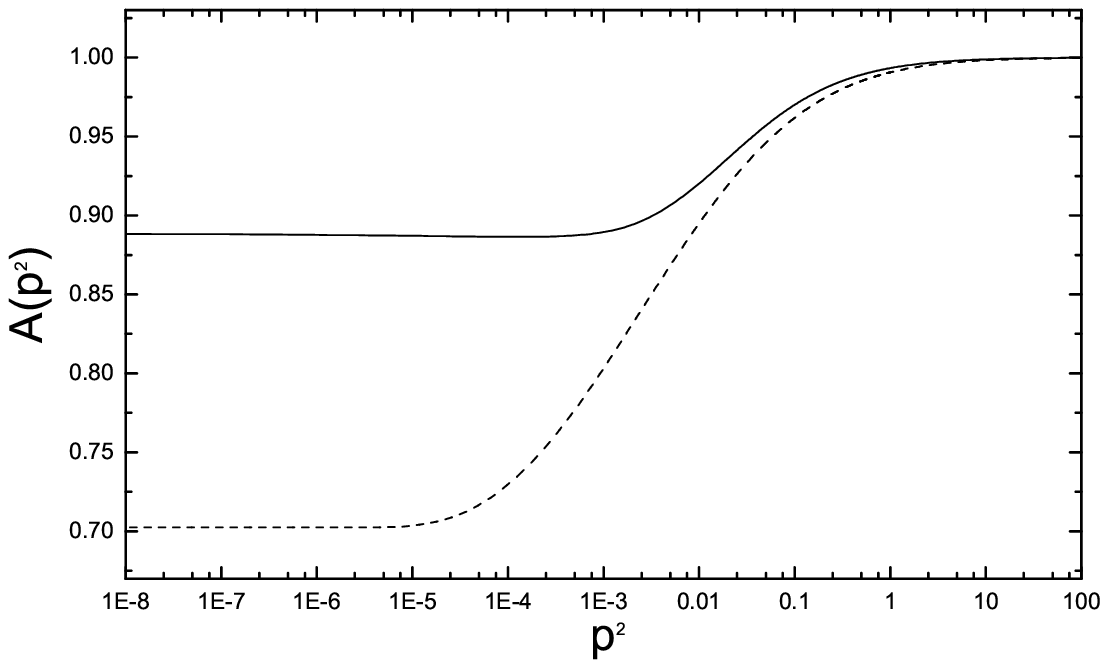}
  \includegraphics[width=0.22\textwidth]{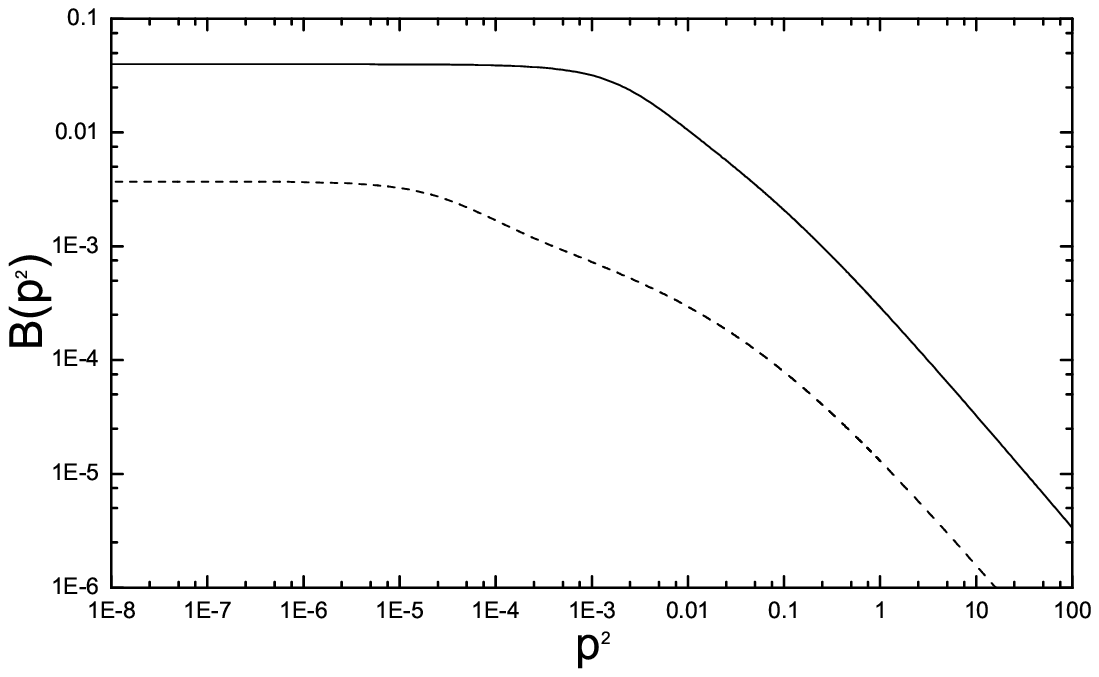}\\
\includegraphics[width=0.22\textwidth]{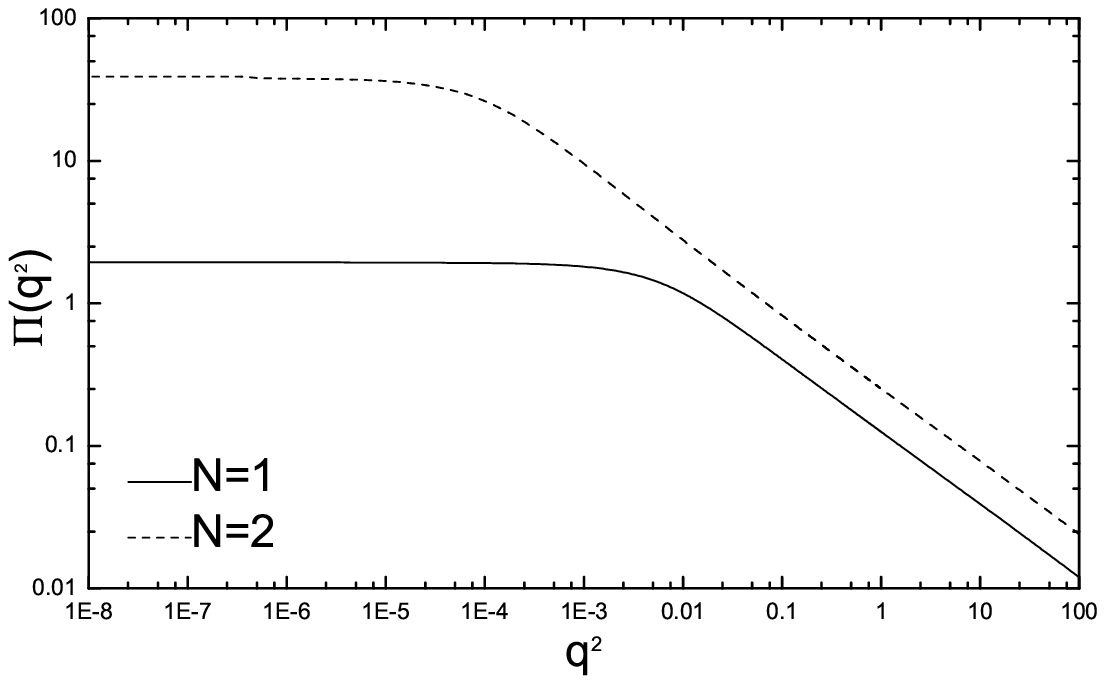}
  \includegraphics[width=0.22\textwidth]{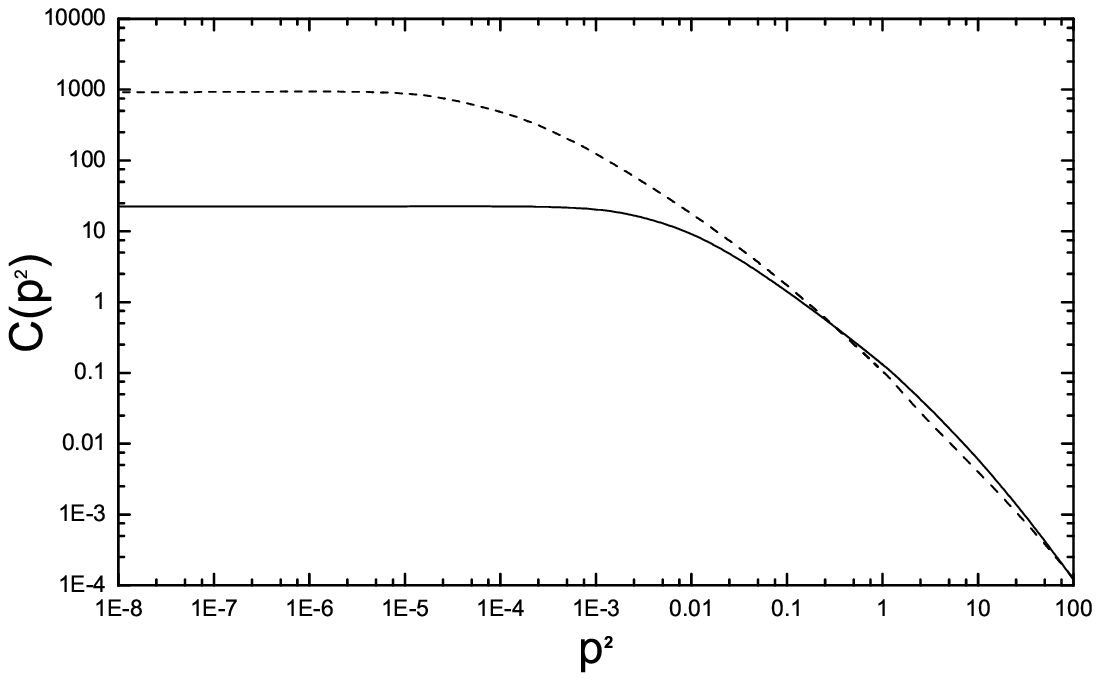}\\
\includegraphics[width=0.22\textwidth]{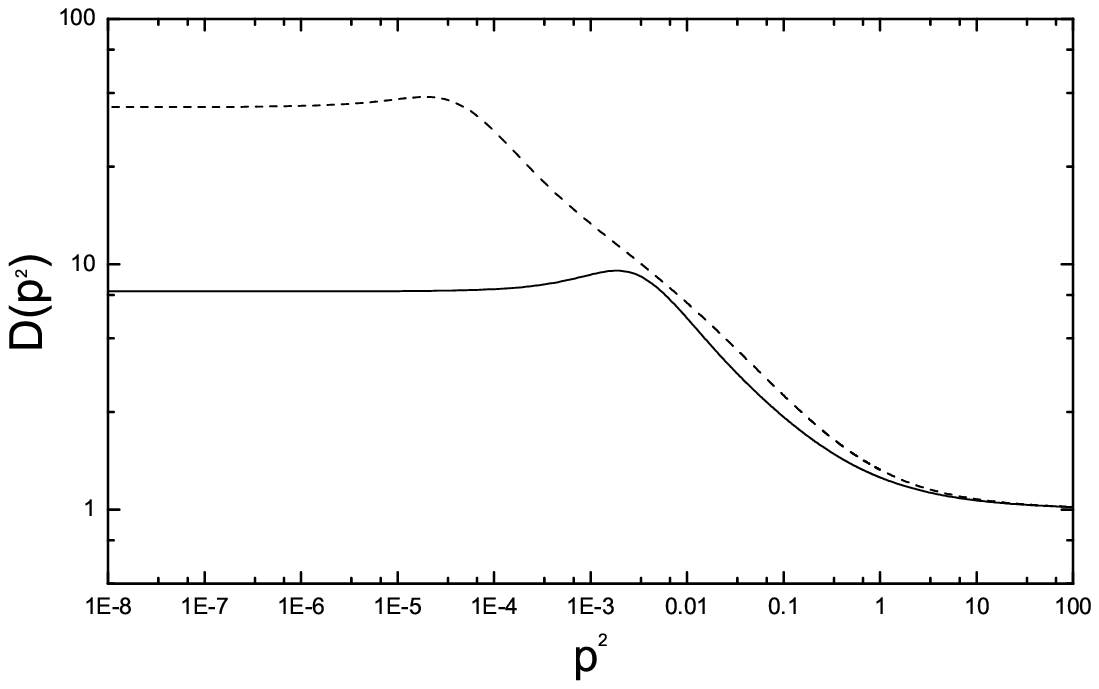}
\includegraphics[width=0.22\textwidth]{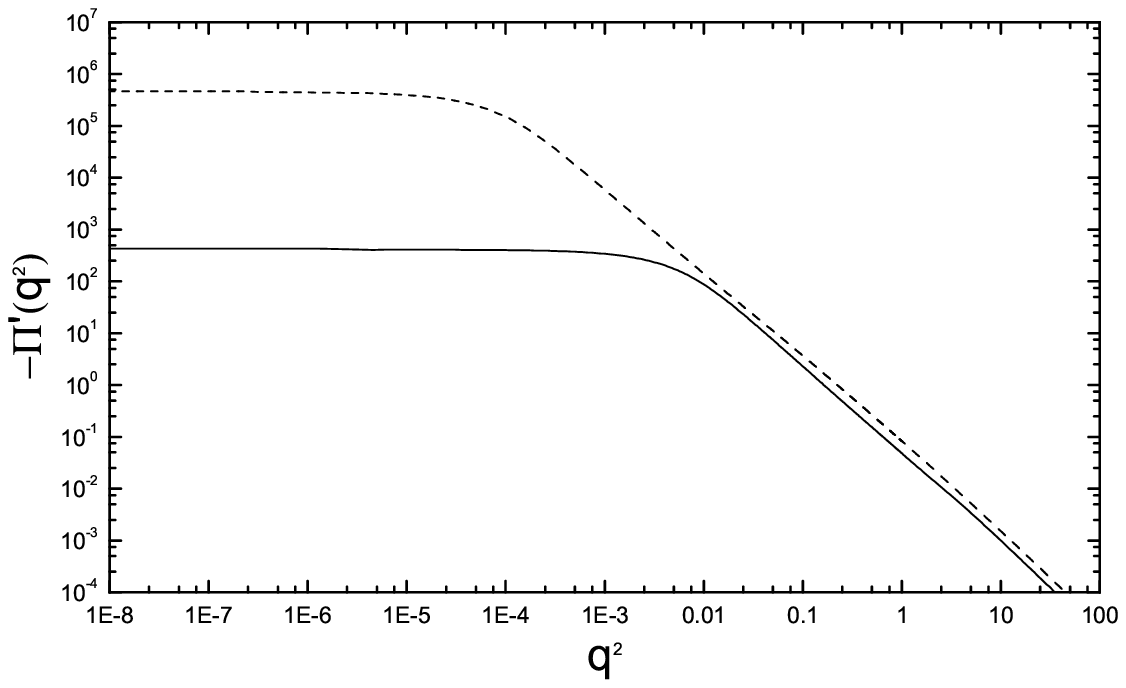}\\
\caption{The behavior of $A(p^2),~B(p^2),~C(p^2),~D(p^2),~\Pi(q^2)$
and $-\Pi'(q^2)$ in BC$_1$ vertex approximation at$N=1, 2$.}\label{FIG1}
\end{figure}

By
application of numerical methods, $A,~B,~\Pi$ and the functions for
the scalar vertex can be obtained. The typical behaviors for the six
functions $A(p^2),~B(p^2),~C(p^2),~D(p^2)$ and $\Pi(q^2),~\Pi'(q^2)$
are shown in Fig. \ref{FIG1}.
From Fig. 1 it can be seen that, excepting that $A(p^2)$ and
$D(p^2)$  approach 1, the other functions vanish in the large
momentum limit and all the six functions are almost constant in the
infrared region.

\begin{figure}[htp!]
  \includegraphics[width=0.52\textwidth]{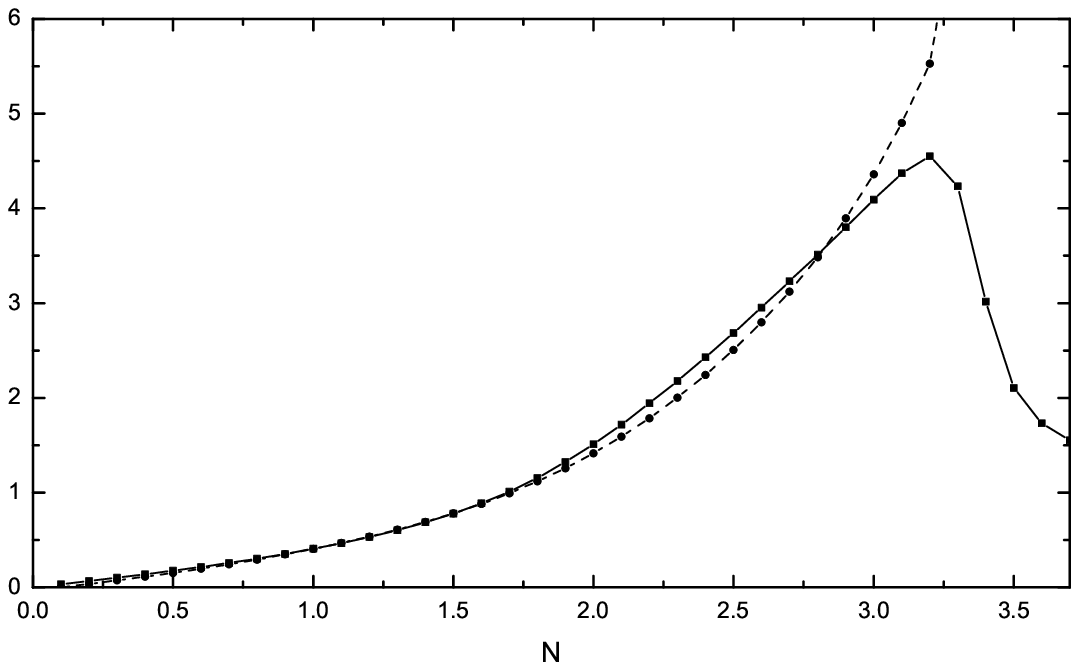}\\[-1cm]
  \includegraphics[width=0.52\textwidth]{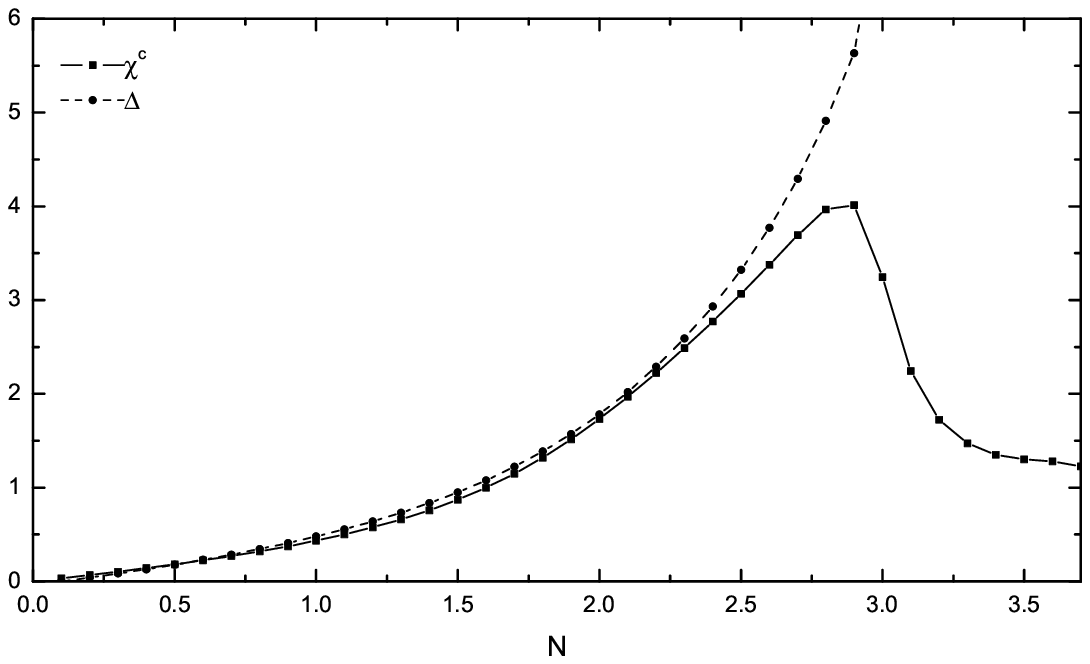}\\[-0.2cm]
  \caption{The dependence of chiral susceptibility and fermion chiral condensate at zero temperature on $N$ in the rainbow approximation (upper pannel) and BC$_1$ vertex approximation
(lower pannel), where
$\Delta=-\lg\frac{\langle\bar\psi\psi\rangle_N}{\langle\bar\psi\psi\rangle_{N=0}}$.}\label{FIG2}
\end{figure}

Substituting the above functions into Eq. (\ref{chic}), we
immediately obtain the value of chiral susceptibility and fermion
chiral condensate with a range of fermion flavors. The results are
plotted in Fig. \ref{FIG2}.
From this figure, we see that, with $N$ increasing, the chiral susceptibility
shows an obvious peak in the rainbow
approximation and BC$_1$ vertex approximation, while
$\langle\bar\psi\psi\rangle$ diminishes and vanishes at a critical
number of fermion flavors where CPT occurs. Since each ansatze
keeps different symmetry of the system, $N_c$ depends a little on
the choice of the ansazte for the dressed vertex. In addition, we
also see that the susceptibility around $N_c$ is apparently
different from that at high temperature and high density \cite{a14}.
The peak shows a neither divergent nor discontinuous behavior which
illustrates that CPT at $N_c$ is neither of first-order nor of second-order
and thus is a higher-order continuous phase transition, which is
consistent with the previous works \cite{a6a,a6b}.

\section{Finite temperature}
With the involved temperature, $O(3)$ symmetry of the system reduce
to $O(2)$ and the gauge boson acquires a nonzero mass. The mass of
the photon implies that external electric fields are screened by
thermal excitations \cite{a7} and so we expect that the feature of
CPT may be changed by the excitations.
\subsection{Truncated DSE}
To give an insight of CPT, we shall adopt DSE for the fermion
propagator and techniques of temperature field theory to calculate
the chiral and thermal  susceptibility  at finite temperature with
the increasing $N$ and analyze the transition of QED$_3$ near
$N_{Tc}$, and also reveal the nature of CPT at the critical
temperature $T_{Nc}$ with a fixed $N$.

Now, let us give a short review of some studies on the effect of the
wave function renormalization factor $A_\parallel$ and $A_3$. Just
as mentioned in Sec. I, the chiral phase transition (CPT) in QED$_3$
was first studied in Ref. \cite{a1}, where it is found that CPT
occurs at $N_c \approx3.24$. They arrived at this conclusion by
solving the lowest order DSE for the fermion self-energy. Later,
some groups adopted improved schemes for DSE to study this problem
and obtained qualitatively similar results with $N_c \approx 3.3$
\cite{a5,a6}. This suggests that the lowest order DSE for the
fermion propagator is a suitable approximation to study CPT at
finite temperature.

At finite temperature, to obtain a qualitative picture of chiral
susceptibility, we employ a familiar framework to obtain the scalar
part of the inverse fermion propagator where the zero frequency
approximation of boson polarization is widely adopted
\cite{a7,a14,c2,c3}. In addition, the conclusions in Ref. \cite{c4}
illustrated that, by summing over the frequency modes and taking
suitable simplifications, the qualitative aspects of the result
obtained under the zero frequency approximation for the wave
function renormalization $A,E$ and the fermion mass function $B,F$
do not undergo significant changes. From this, we also ignore the
frequency dependence of fermion self-energy $B$ and then the
corresponding DSE for the scalar part of inverse fermion propagator
reads \cite{a14}
\begin{eqnarray}\label{fge}
F(P^2)&=&m+2T\int\frac{\mathrm{d}^2K}{(2\pi)^2}\sum_{n=-\infty}^{\infty}\frac{F(K^2)/[Q^2+\Pi(Q)]}{\varpi^2_n+K^2+F^2(K^2)}\nonumber\\
&=&\int\frac{\mathrm{d}^2K}{(2\pi)^2}\frac{F(K^2)\tanh\frac{\mathcal{E}_k}{2T}}{\mathcal{E}_k[Q^2+\Pi(Q)]},
\end{eqnarray}
where $\mathcal{E}_k=\sqrt{K^2+F^2(K^2)}$ and the zero frequency
boson polarization
\begin{equation}\label{PMF}
\Pi(Q)=\frac{NT}{\pi}\int^1_0\mathrm{d}x\left\{\ln\left(4\cosh^2\frac{M(x)}{2T}\right)-\frac{m^2\tanh\frac{M(x)}{2T}}{TM(x)}\right\},
\end{equation} with $M^2(x)=m^2+x(1-x)Q^2$.

With the general equation for the chiral susceptibility
(\ref{chic}), we can obtain the chiral susceptibility at finite
temperature
\begin{eqnarray}\label{CHITD}
\chi^c&=&4NT\sum_n\int\frac{\mathrm{d}^2P}{(2\pi)^2}\nonumber\\&&\times\left\{\frac{[\varpi_n^2+P^2-B^2(P^2)]D(P^2)}{[\varpi_n^2+P^2+B^2(P^2)]^2}-\frac{1}{\varpi_n^2+P^2}\right\}\nonumber\\
&=&2N\int\frac{\mathrm{d}^2P}{(2\pi)^2}\times\nonumber\\&&\left\{\frac{D(P^2)}{\mathcal{E}_p}\left[\frac{P^2}{\mathcal{E}_p^2}\tanh\frac{
\mathcal{E}_p}{2T}\right.+\frac{B^2(P^2)\mathrm{sech}^2\frac{
\mathcal{E}_p}{2T}}{2T\mathcal{E}_p}\right]\nonumber\\&&-\frac{1}{\mathcal{E}_{p0}}\left[\frac{P^2}{\mathcal{E}^2_{p0}}\tanh\frac{
\mathcal{E}_{p0}}{2T}\right\},
\end{eqnarray}
where $\mathcal{E}_{p0}=\sqrt{P^2}.$ The unknown function in the
above equation, $D(P^2)$, is obtained by $F(P^2)$
\begin{eqnarray}\label{CHITF}
&D(P^2)&=\lim_{m\rightarrow0}\frac{\partial F(P^2)}{\partial m}=1+\int\frac{\mathrm{d}^2K}{(2\pi)^2}\frac{1}{\mathcal{E}_k[Q^2+\Pi(Q)]}\times\nonumber\\
&&\left\{\left[\frac{D(K^2)K^2}{\mathcal{E}_k^2}-\frac{B(K^2)\Pi'(Q)}{Q^2+\Pi(Q)}\right]\tanh\frac{\mathcal{E}_k}{2T}\right.\nonumber\\
&&\left.+\frac{D(K^2)B^2(K^2)}{2T\mathcal{E}_k}\mathrm{sech}^2\frac{\mathcal{E}_k}{2T}\right\},\nonumber\\
\end{eqnarray}
with $\Pi'(Q)=\lim_{m\rightarrow0}\frac{\partial \Pi(Q)}{\partial
m}$. From Eq. (\ref{PMF}), we easily find that $\Pi'(Q)=0$.

Similarly, thermal susceptibility measures the response of the
chiral condensate to an infinitesimal change of temperature
\begin{eqnarray}\label{CHIT}
  \chi^T &=& \frac{\partial \langle\bar\psi\psi\rangle}{\partial
  T}\nonumber\\
  &=&\int
\frac{\mathrm{d}^2P}{(2\pi)^2\mathcal{E}_p}\left[B'(P^2)\tanh\frac{\mathcal{E}_p}{2T}-\frac{B^2(P)B'(P)\tanh\frac{\mathcal{E}_p}{2T}}{\mathcal{E}^2_p}\right.\nonumber\\
&&\left.+B(P^2)\mathrm{sech}^2\frac{\mathcal{E}_p}{2T}\left(\frac{B'(P^2)B(P^2)}{2T\mathcal{E}_p}-\frac{\mathcal{E}_p}{2T^2}\right)\right]
\end{eqnarray}
with $B'=\frac{\partial B}{\partial T}$.

\subsection{Numerical results}
\begin{figure}[htp!]
  \includegraphics[width=0.52\textwidth]{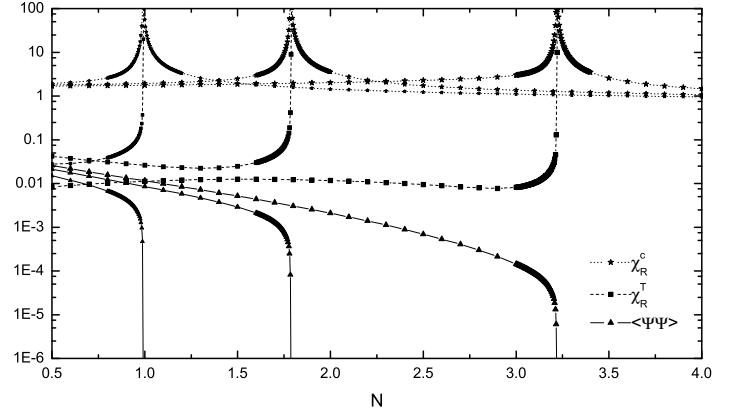}\\
\caption{The behaviors of chiral susceptibility and thermal
susceptibility around the critical fermion flavors with several $T$
(where $\chi^c_R=\frac{\chi^c}{N}, ~\chi^T_R=-\chi^T$, from left to
right denote $T=2.5\times10^{-2}, 10^{-2}, 10^{-3}$).}\label{FIG3}
\end{figure}

\begin{figure}[htp!]
  \includegraphics[width=0.52\textwidth]{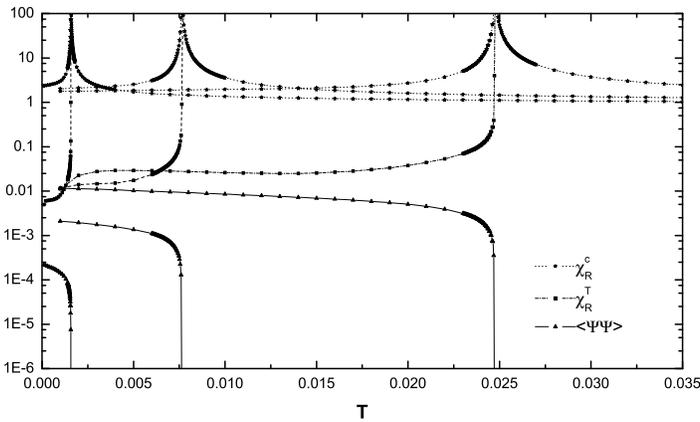}\\
\caption{The behaviors of chiral and thermal susceptibilities around
the critical temperature with several $N$ (from left to right denote
$N=3, 2, 1$).}\label{FIG4}
\end{figure}

After solving the above coupled DSEs by means of the iteration
method, we can now calculate chiral fermion condensate and the above two
susceptibilities, which can be regarded as a function of $N$ given by Eq. (\ref{CHITD})
and (\ref{CHIT}) with a range of temperature. The typical
behaviors of the susceptibilities and condensate are shown in Fig.
\ref{FIG3} and Fig. \ref{FIG4}.
The upper lines of Fig. \ref{FIG3} give the behavior of chiral
susceptibility and the lower lines in this figure show the fermion
chiral condensate, while the other lines between the two group
denote that of thermal susceptibility. As is shown in Fig.
\ref{FIG3}, for any given temperature, $\chi^c$ almost keeps a
constant for small and large $N$, while it shows an apparent peak at
some critical number of fermion flavors. This number depends on the
temperature and diminishes as the temperature increases. When $N$
reaches a critical value $N_{Tc}$, the appearance of vanishing
fermion chiral condensate and divergence peak of $\chi^T$ occur at the same
point. This critical fermion flavors also decreases with the
increase of $T$, which is similar to the results in the previous
works \cite{a7,c2,c3}. By all appearances, at any temperature the
peak of each susceptibility lies at $N_{Tc}$. Moreover, near $N_c$,
the chiral susceptibility at finite temperature shows a different
behavior from that at zero temperature. From Fig. \ref{FIG3}, we
also see that the chiral and thermal susceptibilities exhibits a
very narrow, pronounced and in fact divergent peak at $N_{Tc}$,
which is a typical characteristic of second-order phase transition
driven by the restoration of chiral symmetry at finite temperature.

For a fixed $N$, with the increasing temperature, chiral symmetry
restores at a temperature $T_{Nc}$ and each susceptibility exhibits
the same behavior around the critical point. From Fig. \ref{FIG4},
we see that the chiral and thermal susceptibility reveal their
infinite value at $T_{Nc}$ which also illustrate the typical
second-order phase transition.

\subsection{Critical exponents}
\begin{figure}[htp!]
  \centering
  \includegraphics[width=0.52\textwidth]{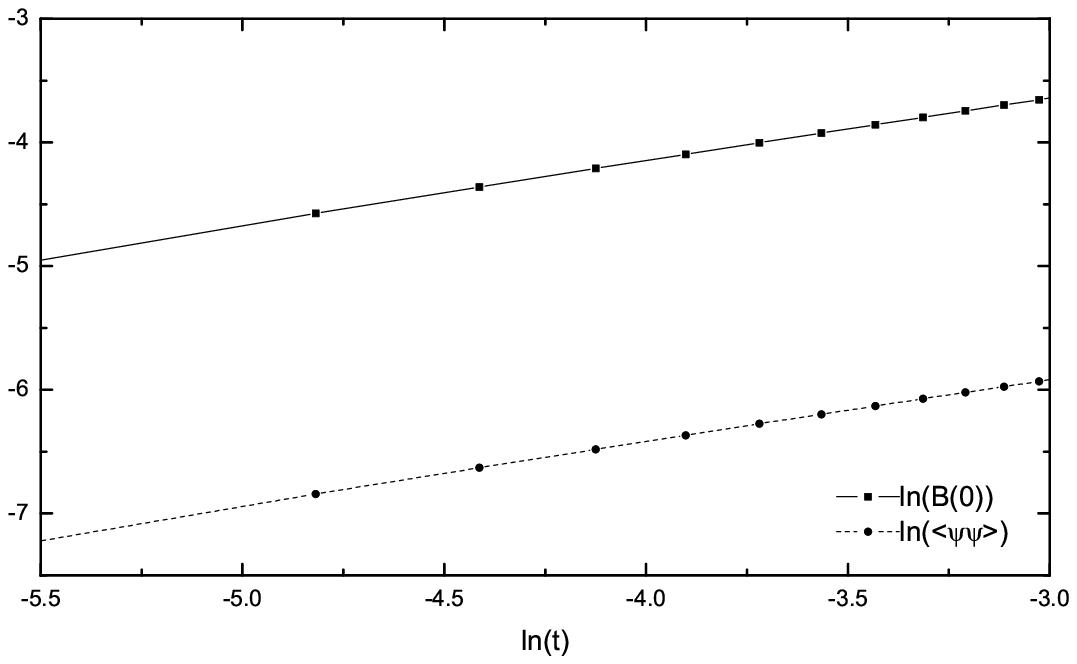}\\[-1cm]
   \includegraphics[width=0.52\textwidth]{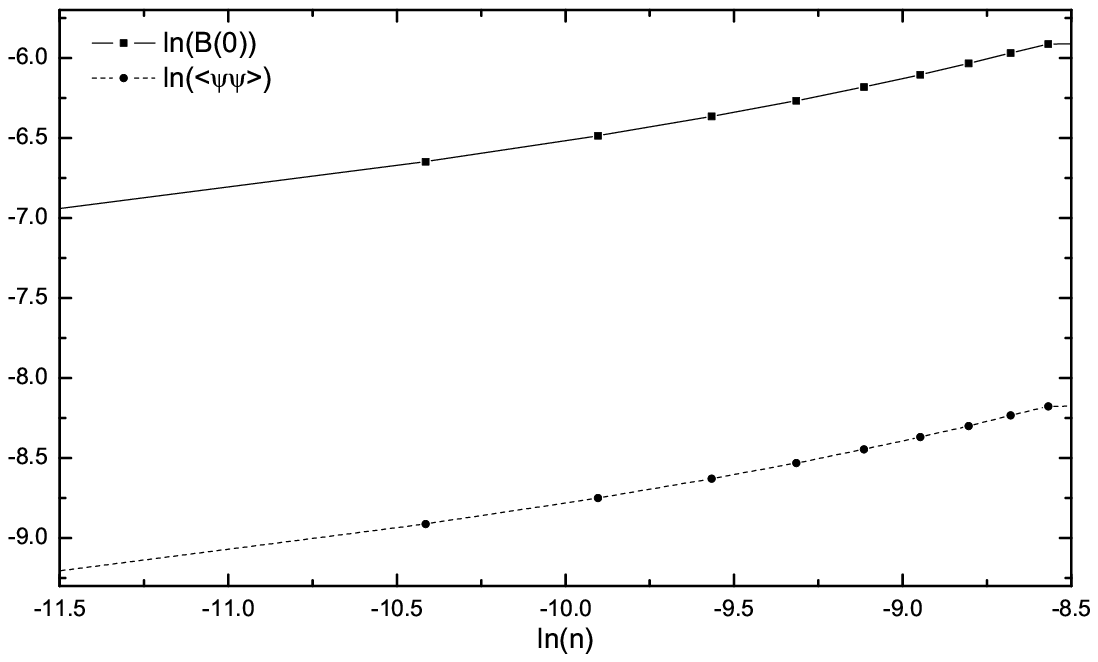}  \\
  \caption{The critical behavior of $\langle\bar\psi\psi\rangle$ and $B(0)$ near the point of CPT with a range of $t$ at $N=1$ (top) and a range of $n$ at $T=0.025$ (bottom). }\label{FIG5}
\end{figure}

Just as shown above, the chiral phase transition at finite
temperature is second order, a natural question is: what are the
critical exponents? Now, let us try to answer this question. Around
the critical points, the phase transitions are characterized by the
corresponding critical exponents which are an important contemporary
goal to exhibit the feature of CPT. We find that the fermion chiral
condensate near the critical point reveals
\begin{eqnarray}\label{CDSBR}
\langle\bar\psi\psi\rangle\sim t^\alpha,~N=const,\nonumber\\
\langle\bar\psi\psi\rangle\sim  n^\beta,~T=const,
\end{eqnarray}
with the reduced temperature $t=1-T/T_c$ and the reduced fermion flavors number $n=1-N/N_c$. The typical behavior of the
condensate near the point of CPT can be seen in Fig. \ref{FIG5} and find
that, in each figure, the slope of the line of infrared fermion
self-energy $B(0)$ is same to that of $\langle\bar\psi\psi\rangle$
which indicates that $B(0)$ and $\langle\bar\psi\psi\rangle$
illustrate the same value of critical exponent in massless QED$_3$.

\begin{figure}[htp!]
  \centering
  \includegraphics[width=0.52\textwidth]{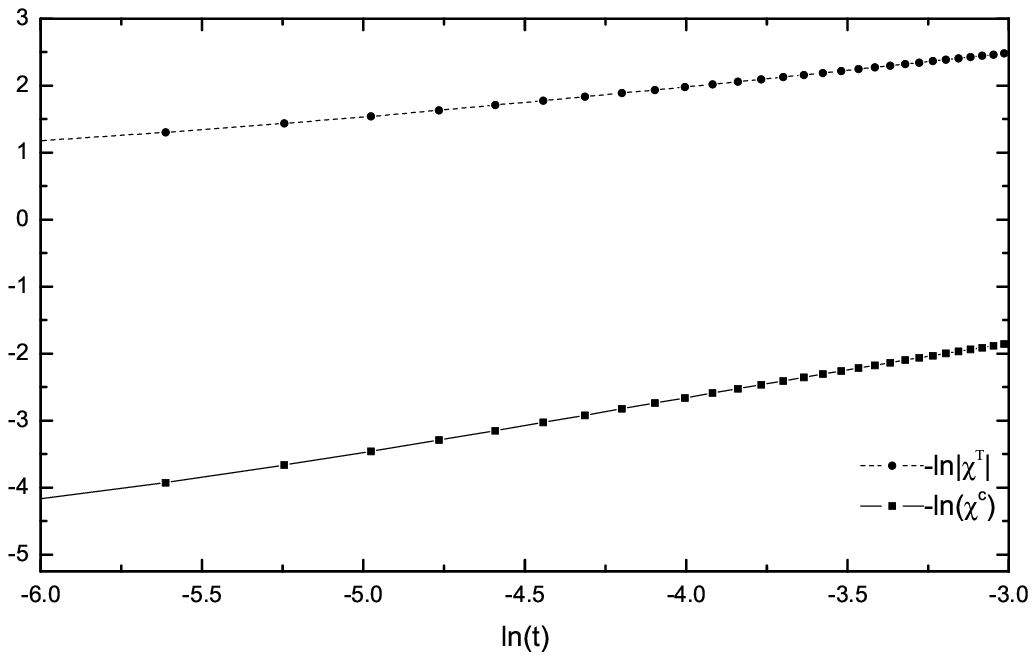}\\[-1cm]
   \includegraphics[width=0.52\textwidth]{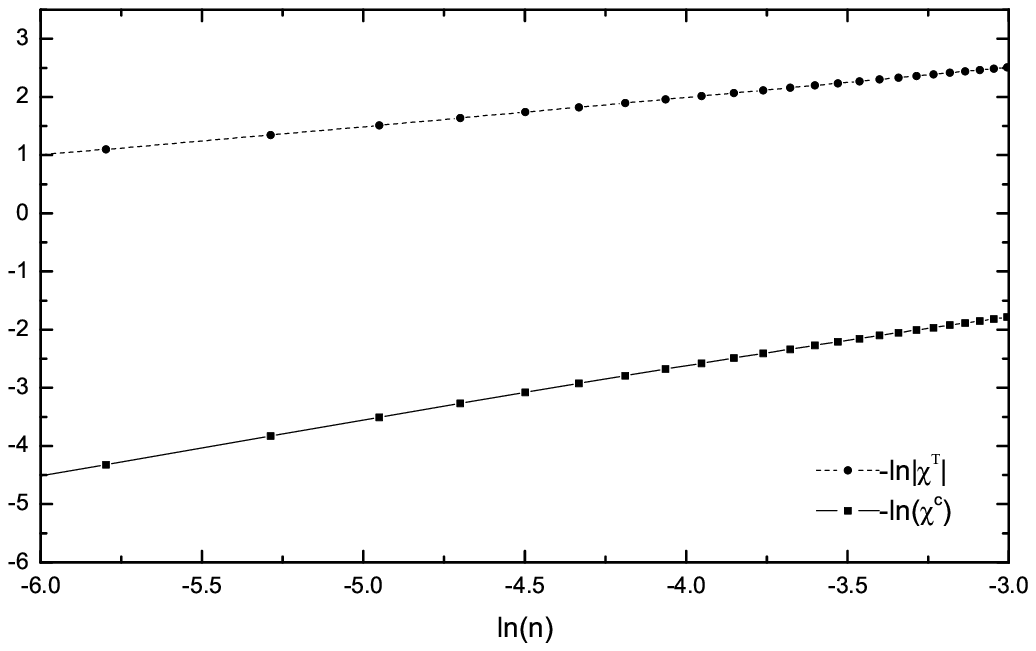}  \\
  \caption{The critical behavior of two susceptibilities near the point of CPT with a range of $t$ at $N=1$ (top) and a range of $n$ at $T=0.025$ (bottom).}\label{FIGEND}
\end{figure}

Numerically, for a fixed $N$, the estimated $\alpha$  at
$t\rightarrow 0^+$ and also the estimated $\beta$ at $n\rightarrow
0^+$ with several $T$ are given as
\begin{center}
\begin{tabular}{c|c||c|c}
  \hline \hline
 N &    $\alpha$ &T & $\beta$   \\
 \hline
1&  0.507& $10^{-3}$ &0.487\\
2&  0.538& $10^{-2} $& 0.401 \\
3&  0.416 & 0.025  & 0.423\\
  \hline
\end{tabular}
\end{center}
In addition, near the point of phase transition,  each of the two
susceptibilities  at $t, n\rightarrow0^+$ reveals its critical
feature as
\begin{eqnarray*}\label{CRI}
    \chi^{c}\sim t^{-\gamma^{c}},~N=const,\\
    \chi^{T}\sim t^{-\gamma^{T}},~N=const,\\
   \chi^{c}\sim n^{-\delta^{c}},~T=const,\\
    \chi^{T}\sim n^{-\delta^{T}},~T=const,
\end{eqnarray*}
and their critical behaviors can be found in Fig. \ref{FIGEND}.

From the numerical results, we estimate the critical exponents of
the susceptibility with several $N$ or $T$ and give $\gamma,~\delta$
in the following table:
\begin{center}
\begin{tabular}{c|c|c||c|c|c}
  \hline
    \hline
  N & $\gamma^c$ & $\gamma^T$  & T & $\delta^c$  &$\delta^T$  \\
  \hline
 1 & 0.679 & 0.350 & $10^{-3}$ & 0.769 & 0.622 \\
  2 & 0.769 & 0.354& $10^{-2}$ & 0.813 & 0.712 \\
  3 & 0.476 & 0.274 & 0.025 & 0.931& 0.455 \\
  \hline
\end{tabular}
\end{center}

It is shown that each of the critical exponents is less than 1 and,
in the same boundary condition, the critical exponent of $\chi^c$ is
larger than that of $\chi^T$.

\section{conclusions}
The primary goal of this paper is to investigate the nature of
chiral phase transition of QED$_3$ near the critical value,
including critical number of fermion flavors and critical
temperature through a continuum study of the chiral and thermal
susceptibilities. Based on the suitable approximation of truncated
DSEs for the fermion propagator and numerical model calculations, we
study the behavior of the two susceptibilities near the critical
point of CPT in QED$_3$. It is found that, with the rise of the
number of fermion flavors, the appearance of the peak of chiral
susceptibility and CPT occur at the same critical point, but the
peak reveals apparently different behavior at zero and finite
temperature.

At zero temperature the chiral susceptibility near the critical
number of fermion flavors reveals a finite and continuous peak,
which exhibits that CPT is neither of first order nor of second
order, and thus it should be a continuous phase transition of higher
order. However, apart from zero temperature, each of chiral and
thermal susceptibility at either critical fermion flavors or chiral
temperature shows a large and in face divergent peak which
illustrates a typical characteristic of second-order phase
transition driven by chiral symmetry restoration in thermal QED$_3$.

Finally, though the analysis for the critical exponents, it is found
that the critical exponents of chiral/thermal susceptibility which
characterize the chiral phase transition is between 0.2 and 1 and,
in the same boundary condition, the critical exponent of thermal
susceptibility is less than that of chiral susceptibility.

\section{acknowledgements}

This work was supported in part by the National Natural Science
Foundation of China (under Grant Nos. 11105029, 11275097, and
11347212) and the Research Fund for the Doctoral Program of Higher
Education (under Grant No 2012009111002) and by the Fundamental
Research Funds for the Central Universities (under Grant No
2242014R30011).

\end{document}